\documentclass[10pt]{wlscirep}
\usepackage[utf8]{inputenc}
\usepackage[T1]{fontenc}

\usepackage{multirow}

\usepackage{calrsfs}
\DeclareMathAlphabet{\mathcal}{OMS}{zplm}{m}{n}

\title{Hidden transition in multiplex networks}

\author[1,*]{R.~A.~da~Costa}
\author[1]{G.~J.~Baxter}
\author[1]{S.~N. Dorogovtsev}
\author[1]{J.~F.~F. Mendes}

\affil[1]{Department of Physics {\&} I3N, University of Aveiro, Portugal}

\affil[*]{americo.costa@ua.pt}

\begin{abstract}
Weak multiplex percolation generalizes percolation to multi-layer networks, represented as networks with a common set of nodes linked by multiple types (colors) of edges.
We report a novel discontinuous phase transition in this problem. 
This anomalous transition 
occurs in networks of three or more layers without unconnected nodes, $P(0)=0$. 
Above a critical value of a control parameter,
the removal of a tiny fraction $\Delta$ of nodes or edges triggers a failure cascade which ends either with the total collapse of the network, or a return to stability with the system essentially intact.
The discontinuity is not accompanied by any singularity of the giant component, in contrast to the discontinuous hybrid transition which usually appears in such problems.
The control parameter is the fraction of nodes in each layer with a single connection, $\Pi=P(1)$. We obtain asymptotic expressions for the collapse time and relaxation time, above and below the critical point $\Pi_c$, respectively.
In the limit $\Delta\to0$ the total collapse for $\Pi>\Pi_\text{c}$ takes a time $T \propto 1/(\Pi-\Pi_\text{c})$, while there is an exponential relaxation below $\Pi_\text{c}$ with a relaxation time $\tau \propto 1/[\Pi_\text{c}-\Pi]$.
\end{abstract}
\begin{document}

\flushbottom
\maketitle

\thispagestyle{empty}

\section*{Introduction}
\label{intro}

The interest in multilayer networks has been growing in the past few years due to both their ability to describe higher levels of complexity present in many real systems, and to the new and surprising properties that emerge from the interaction among layers\cite{buldyrev2010catastrophic,gao2011robustness,kivela2014multilayer,boccaletti2014structure,bianconi2018multilayer}. 
In many real networks there are different kinds of dependencies and relations between nodes, including infrastructure \cite{rinaldi2001identifying}, information \cite{leicht2009percolation}, financial \cite{caccioli2014stability, huang2013cascading} ecological \cite{pocock2012robustness} , and many other systems \cite{min2014multiple,duenas2007seismic,bianconi2016percolation}. One may represent these systems as multilayer networks that have a layer for each kind of relation (edge color) and where nodes are present in all layers.
These can be regarded as multiplex (or colored) networks, that is, networks containing edges of multiple colors corresponding to multiple kinds of relations \cite{son2012percolation}.

Similarly to single layer (uniplex) networks, the effects of local failures in these networks are described by their percolation properties. 
However, the generalization of percolation to multiplex networks is not unique. 
Mainly, there are two alternative definitions generalizing the notion of cluster in multiplex percolation: the (i) mutually connected cluster, which relies in a global criterion, and the (ii) weakly percolating cluster which is defined through a local, more permissive, rule \cite{baxter2016unified,baxter2021weak}.
The stronger definition of (i) requires that each pair of nodes in the same cluster is connected by at least one path inside of each of the layers. 
This rule is a global one in the sense that it explicitly requires us to verify large-scale (global) connectivity, namely to find the giant mutually connected cluster \cite{buldyrev2010catastrophic}. 
The weaker definition of (ii) only requires that every node in a cluster has at least one connection in each layer to another node in the same cluster \cite{baxter2014weak}. 
With this local rule the survival of a node in a cluster depends strictly on its neighbors.
In networks of a single layer both rules reduce to the standard definition of cluster.

\begin{figure}[h]
\begin{center}
\includegraphics[scale=0.65]{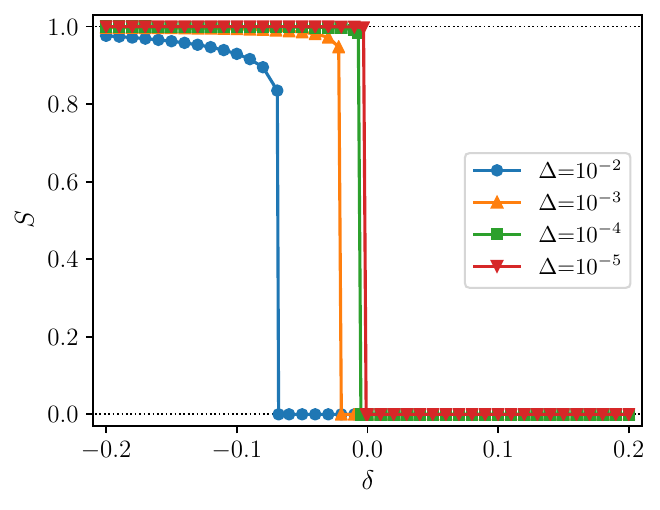}
\end{center}
\caption{
Size of the giant weak percolation component $S$ after the cascade of failures following the removal of edges with probability $\Delta$ in simulations of uncorrelated networks with $N=10^8$ and $m=3$ layers of edges, as a function of reduced control parameter $\delta = \Pi - \Pi_c$, where $\Pi=P(1)$. In the limit of very small damage, the behavior converges to a novel discontinuous transition. 
The degree distribution is $P(0){=}0$ and $P(q){=}e^{-c}c^{q-1}/(q{-}1)!$ for $q{>}0$, with $c$ such that $P(1){=}e^{-c}{=}1/2 {+} \delta$.
Each point is averaged over $10$ realizations of the network and damage. Lines are simply a guide to the eye.
}
\label{fig_S_vs_P1}       
\end{figure}

Despite the different characters of these two definitions of cluster, they have many similar effects in the percolation transition.
In particular, both generalizations of the percolation problem for multilayer networks typically display hybrid phase transitions, with characteristics from both continuous and discontinuous transitions \cite{buldyrev2010catastrophic, baxter2012avalanche, baxter2021weak}, of the same type as observed in $k$-core percolation \cite{dgm2006}.
This transition persists even in correlated multiplex networks \cite{hu2013percolation,min2015link,baxter2016correlated,cellai2016message}.
The order parameter, which is traditionally regarded as the size of the giant component, jumps from zero to some finite value at the critical point in a discontinuous transition.
This jump is followed by a critical square root singularity in the ordered phase, associated with a diverging susceptibility (defined as mean avalanche size induced by the removal of a random node or edge) as in continuous transitions,  and diverging relaxation times \cite{baxter2015critical}.
This hybrid character of the transition is typically found in networks with two or more layers in the case of the stronger definition of cluster (i), while with the weaker definition (ii) at least three layers are required, otherwise the transition is continuous.
Exceptions are found in some special cases, as, for example, in networks with power-law degree distributions with very low exponents, which display continuous growth of the giant cluster even  with a large number of layers \cite{baxter2020exotic}.
Note that the weaker definition provides a particular case of the ${\bf k}$-core problem for multiplex networks \cite{azimi2014k}, for the threshold ${\bf k} = (1,1,\ldots,1)$, and a similar model was later proposed in Ref. \cite{dimuro2017cascading}.

Here we show that the weakly percolating cluster also exhibits a novel discontinuous transition that is not observed in other multilayer network problems. A weakly percolating cluster, by definition, contains no nodes with degree zero, $P(0)=0$ in all layers. Starting from a stable giant weakly percolating component, a small amount of random damage reveals the hidden discontinuous transition, see Fig.~\ref{fig_S_vs_P1}. Interestingly, the critical point is defined by the fraction of nodes of degree $1$, which we use as the control parameter.
Above the transition threshold the giant component collapses completely under perturbation, while above the transition it remains intact. The transition is discontinuous without any precursor, unlike the hybrid transition which usually appears in such systems. Discontinuous transitions without any critical exponent have been observed in some models of dynamic processes or with spatially limited connections in multilayer networks \cite{zhang2018cascading,danziger2014percolation,danziger2015interdependent}.

We characterise this hidden transition, and give conditions for its appearance. We further examine the dynamics of the collapse, showing that it is associated with long lived relaxation both above and below the transition, and which diverge approaching the critical point. The form of the divergence depends in a complex way on either the distance from the critical point or the size of the perturbation, depending on the relation between them. This behavior is quite different from that observed in the usual hybrid transition.


\subsection*{Weak percolation on $m$-layer multiplex}
\label{s1} 

Let us consider a multiplex network with $m$ layers, or colors of edges.
For the sake of simplicity we will consider random networks without degree-degree correlations within individual layers.
Furthermore, our analysis will mainly focus on networks also without correlations between degrees of the same node in different layers, however, for illustrative purposes, we briefly consider networks containing an extreme form of degree correlations. 
That extreme example of layer degree correlations allows to show that their presence does not change the nature of the phenomena.

The standard method for finding weakly connected components in particular realizations of multiplex networks consists of a simple pruning procedure. 
At each pruning step, we keep the nodes that have at least one surviving neighbor in every layer, and remove all nodes that fail to meet that criterion.
The process stops when all of the surviving nodes meet this criterion, and the remaining network is formed only of weakly connected clusters.
In locally tree-like random networks, finite weakly connected clusters cannot exist, thus, the only possibility for a cluster to survive this pruning process is to be the giant cluster \cite{baxter2014weak}.

An $m$-layered system is fully described by the degree distribution $P(q_1, ... ,q_m)$ of a randomly selected node having degrees $q_1$ in layer $1$, $q_2$ in layer $2$, etc.
For locally tree-like networks, the weak percolation problem is expressed by the following system of $m$ self-consistency equations
\begin{equation}
Z_\alpha = \sum_{q_1, ... ,q_m} \frac{q_\alpha P(q_1, ... ,q_m)}{\langle q_\alpha \rangle} \prod_{\beta\neq\alpha} \bigl[1-(1-Z_\beta)^{q_\beta}\bigr]
,
\label{10}
\end{equation}
where
$\alpha=1, ... , m$, and $Z_\alpha$ is the probability that following a random edge in layer $\alpha$ leads to a node which has at least one edge in layer $\beta$ satisfying the equivalent condition, which occurs with probability $Z_\beta$, for all layers $\beta\neq\alpha$.
The relative size of the giant cluster may then be determined by inserting the solution of Eq.~(\ref{10}) in the following expression,
\begin{equation}
S = \sum_{q_1, ... ,q_m}  P(q_1, ... ,q_m) \prod_{\beta=1}^m \bigl[1-(1-Z_\beta)^{q_\beta}\bigr]
.
\label{20}
\end{equation}

The solution of the self-consistency system can be found by iterating Eq.~(\ref{10}).
For convenience let us denote by $Z_\alpha(t)$ the value obtained for $Z_\alpha$ at iteration $t$.
We start from $Z_\alpha(0)=1, \forall \alpha$, and calculate the values of $Z_\alpha(t+1)$ by inserting the values of the previous iteration,  $Z_\beta(t)$, on the right-hand side of Eq.~(\ref{10}). 
The sequence of $Z_\alpha(t)$ generated in this way is guaranteed to converge to the solution of the self-consistency equations at large enough $t$.

In fact, this iterative process is closely related to the pruning procedure used to extract the weakly connected components described above.
Each value of the sequence ${Z_\alpha}(t)$ is the probability that following a random edge in layer $\alpha$ of the original network leads to a node still remaining at the $t$-th pruning step.
Similarly, inserting the values ${Z_\alpha}(t)$ in the right-hand side of Eq.~(\ref{20}) gives the fraction of nodes still remaining at the $t$-th pruning step $S(t)$.
Notice that despite being written for the ensemble averages of infinite random networks generated by the distribution $P(q_1,...,q_m)$,  Eqs.~(\ref{10}) and~(\ref{20}) are also valid for individual members of the ensemble, due to self-averaging in infinite networks.
Therefore, it is safe to expect that the properties of large, yet finite, networks closely resemble those of their infinite counterparts.

When every node of a network has at least one connection in each layer, i.e. if $P(q_1,...q_m)$ equals zero whenever some $q_\alpha$ is zero, Eq.~(\ref{10}) has a solution at $Z_1=...=Z_m=1$. 
In this case, by definition, every node meets the local criterion of weak percolation; consequently the pruning procedure removes no nodes at all, and, correspondingly, $Z_1=...=Z_m=1$ is a fixed point of the iterative solution of Eqs.~(\ref{10}).
However, this fixed point can be stable or unstable depending on the structure of the network.
As we show in the following, in fact, there is a non-trivial truly discontinuous percolation transition that is caused by the change in the stability of that fixed point.


\section*{Results}


\subsection*{Revealing the hidden transition in weak percolation}
\label{s2}

\begin{figure*}[htp]
\begin{center}
\includegraphics[scale=0.75]{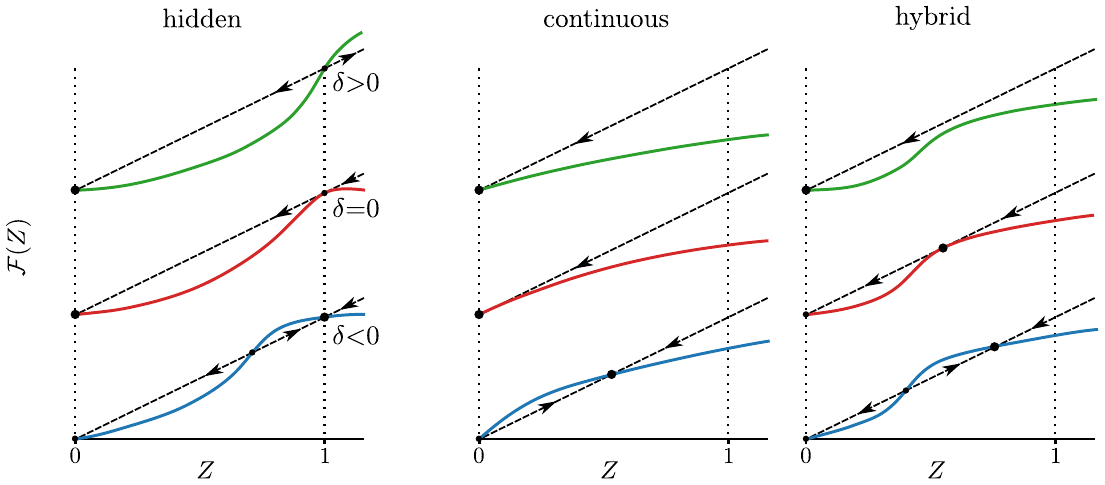}
\end{center}
\caption{
Qualitative behavior of the function ${\cal F}(Z)$ in $m\geq 3$ layers, (leftmost panel). From top to bottom the function is shown above (collapse), at, and below (stable) the point of the hidden transition, respectively.
The fixed points occur when the curve crosses the line ${\cal F}(Z)=Z$, indicated by the dashed line, i.e. when the value of the function equals $Z$.
The stability of the fixed points is indicated by the arrows.
An undamaged multiplex network with a hidden transition stays in the state $Z=1$ above, at, and below the transition point, but any damage results in the final state $Z=0$ at and below the transition ($\delta \geq 0$). 
For comparison, we illustrate the behavior of the corresponding function for the usual hybrid transition (middle panel) and a continuous transition (right panel).
}

\label{f2}       
\end{figure*}

Requiring only that $P(0)=0$, multiplex networks undergo an unusual truly discontinuous (not hybrid) percolation transition.
This transition will remain hidden without perturbations, but can be revealed by application of a tiny amount of damage.
The control parameter is simply the fraction of nodes of degree $1$; the critical point does not depend on the rest of the degree distribution. 
On one hand, when the fixed point $Z_1=...=Z_m=1$ is unstable, even a tiny fraction of removals is enough to start a large-scale cascade of failures, eventually collapsing the whole system.
On the other hand, when the fixed point is stable, a small fraction of removals of nodes or edges will leave the system essentially intact.

For simplicity, let us first consider the symmetric uncorrelated weak percolation problem, that is, $P(q_1,q_2,...,q_m)=P(q_1)P(q_2)...P(q_m)$, where $P(q)$ is any degree distribution with $P(0)=0$.
The self-consistency system of Eq.~(\ref{10}) reduces to a single equation, 
\begin{equation}
Z = \left\{\sum_q P(q) \left[1-(1-Z)^q\right] \right\}^{m-1} \equiv {\cal F} (Z)
,
\label{2300}
\end{equation}
where $Z$ stands for the value of the $Z_\alpha$'s, which, in this case, are all the same. Note that we may now complete the sum over $q_{\alpha}$ in Eq.~(\ref{10}), so that the factor $q_{\alpha}/\langle q_\alpha \rangle$ disappears.

To analyze the stability of the solution $Z=1$ it is convenient to expand Eq.~(\ref{2300}) around that point.
In the iterative form $Z(t{+}1)={\cal F}[Z(t)]$ we get
\begin{equation}
Z(t{+}1)= 1-{\cal F}'(1) [1{-}Z(t)]+{\cal F}''(1) [1{-}Z(t)]^2+...\,.
\label{2400}
\end{equation}
Notice that ${\cal F}(1){=}1$ only if $P(0){=}0$.
The coefficient of the second term in the right-hand side of Eq.~(\ref{2400}), ${\cal F}'(1) {=} (m{-}1)P(1)$, determines the stability of the fixed point $Z{=}1$.
When $1{-}Z(t)$ is close to $0$, if ${\cal F}'(1){>} 1$ then $Z(t{+}1){<}Z(t)$ and (as long as $P(2)$ is not too small, see below) the sequence runs away from the fixed point.
If ${\cal F}'(1){<} 1$ then $Z(t{+}1){>}Z(t)$ and the sequence converges to the fixed point $Z{=}1$.

Since the stability transition occurs when $(m{-}1)P(1){=}1$, we may use $\Pi \equiv P(1)$ as a control parameter.
Thus, the critical point of this transition in uncorrelated networks is 
\begin{equation}
\Pi_\text{c}=1/(m{-}1).
\label{2410}
\end{equation}
The stable phase corresponds to values of $\Pi{<}1/(m{-}1)$.
In the unstable phase, for $\Pi{>}1/(m{-}1)$, a small amount of damage will trigger a large-scale cascade of failures, collapsing the whole system.
Interestingly, networks with no more than two layers have $\Pi_\text{c}{=}1$ and are always stable under small perturbations, while networks with more than two layers can be stable or unstable depending of $\Pi$, since for those $\Pi_\text{c}{<}1$. 
In Methods 
 we give more detail of the stability analysis in two, and three or more layers.
Fig.~\ref{f2} shows typical profiles of the function ${\cal F}(Z)$ 
for $m\geq 3$ above, below, and at the critical point $\Pi_\text{c}$, indicating also the stability of the fixed points in these regimes. For comparison, we illustrate the corresponding functions for the hybrid transition, which occurs in weak multiplex percolation in $m\geq3$ layers, and the continuous transition which occurs in $m=2$ layers, and in ordinary percolation in one layer.

The size of the giant component in the network with infinitesimally small damage is 
\begin{equation}
S = \begin{cases}
1 & \mbox{if } \Pi < \Pi_\text{c},
\\
0 & \mbox{if }  \Pi > \Pi_\text{c}.
\label{e2}
\end{cases}
\end{equation}
Note that this transition is not hybrid (for $m\geq3$). There is a discontinuity without any singularity, the jump is from $1$ to $0$. 
On the other hand, it occurs just at the stability limit, similarly to the hybrid phase transitions.
Fig.~\ref{fig_S_vs_P1} shows the size of the giant weak percolation component at the end of cascade of removals triggered by the application of damage in simulations of large networks, with $N=10^8$ nodes, with $P(q){=}e^{-c}c^{q-1}/(q{-}1)!$, for $q >0$, i.e. a Poisson distribution in $q-1$, with $P(0) = 0$. In the limit of very small damage, the behavior converges to that described by Eq.~(\ref{e2}).

What happens exactly at $\Pi=\Pi_\text{c}$ depends on the sign of the third term in the right-hand side of Eq.~(\ref{2400}). 
If ${\cal F}''(1)$ is positive the fixed point will be stable and if it is negative the fixed point is unstable. 
Near the critical point this problem shows a rich variety of different behaviors, which will be explored in detail in the next section, 
devoted to the dynamics of the collapse process.

The presence of correlations between layer degrees does not change the nature of the hidden weak percolation transition.
In Methods 
 we describe the illustrative example of ultimate correlations, where each node has exactly the same degree in every layer.
All of the results in that situation are similar to the ones for uncorrelated networks. 
The main difference is the position of the critical point, which will be determined by the slightly different condition $(m{-}1)P(1)/\langle q \rangle=1$.
Note that, also in these ultimately correlated networks, a stability transition can only exist if the number of layers $m \geq 3$, and for $m=2$ the system is always stable.

This kind of transition occurs only for the definition of cluster of weak percolation, and not for the mutual connected cluster.
For symmetric uncorrelated networks, the self-consistency equation for the mutual connected cluster is
\begin{align}
Z = \!\left\{ \! \sum_q\! P(q)\! \left[1{-}(1{-}Z)^q\right] \!  \right\}^{\!m-1} \! \sum_q \! \frac{q P(q)}{\langle q \rangle}\! \left[1{-}(1{-}Z)^{q-1}\right]
\equiv {\cal F}_{\text{MCC}}(Z)
.
\label{2301}
\end{align}
This equation also has a solution at $Z=1$ when $P(0)=P(1)=0$, but the fixed point is always stable.
In fact, the limit of stability, ${\cal F}'_{\text{MCC}}(1)=1$, in this case is reached only when $P(2)=1$ for any $m$, and an unstable phase does not exist in that problem.


\subsection*{Dynamics of the collapse}
\label{s3}

Let us consider the effects that the application of damage has in the integrity of the weak percolation component.
When every node has at least one connection in every layer, i.e., $P(0)=0$, the pruning process cannot remove any of the nodes, hence the fixed point $Z=1$.
However, the deletion of even a fraction of the edges or nodes from the original network configuration will cause a few nodes to effectively lose all connections in one of the layers, allowing the pruning process to remove them and starting a cascade of failures.

Let us introduce a probability $\Delta$ that an edge is deleted from the initial configuration.
Then, for uncorrelated networks, the iterative form of Eq. (\ref{2300}) becomes 
\begin{equation}
Z(t+1)=(1{-}\Delta){\cal F}[Z(t)]\,.
\label{2500}
\end{equation} 
Notice that, when we delete nodes with probability $\Delta$, instead of edges, the self-consistency equation is the same as Eq.~(\ref{2500}), the difference between the two deletion schemes will be only in the equation for $S$. 
So, the following results apply to node deletion as well.

For convenience, let us write the expansion of Eq.~(\ref{2500}) in terms of the quantity $\zeta{=}1{-}Z$, 
\begin{equation}
\zeta(t+1) = \Delta + (1{-}\Delta)[{\cal F}'(1) \zeta(t) - {\cal F}''(1) \zeta^2(t) +...] .
\label{2600}
\end{equation}
Equation~(\ref{2600}) provides an accurate description of the evolution of $Z=1-\zeta$ during the pruning process as long as $\zeta$ is small.

\begin{figure}[h]
\begin{center}
\includegraphics[scale=0.5]{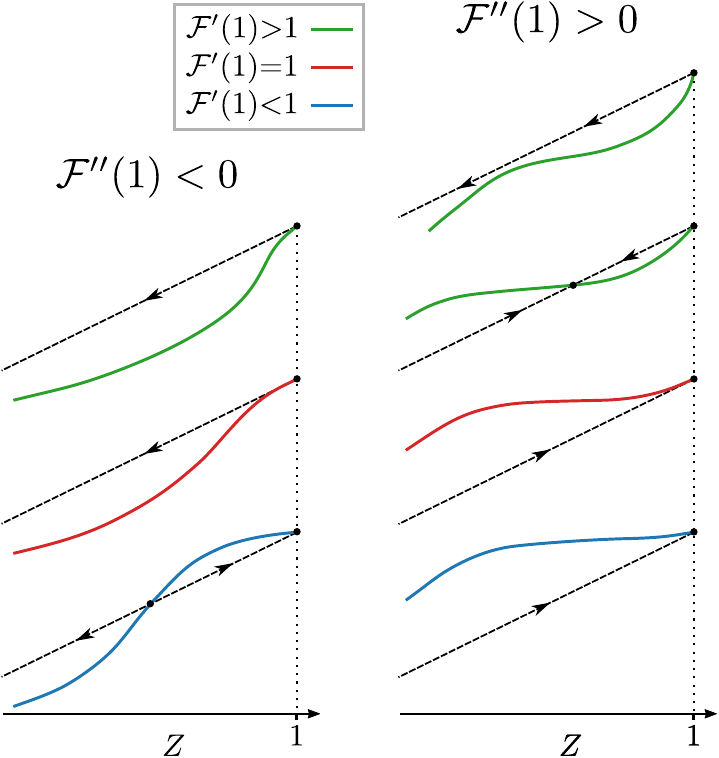}
\end{center}
\caption{
Schematic representation of function ${\cal F}(Z)$ [see Eq.~(\ref{2300})] near $Z=1$, in two different scenarios depending on the sign of the second derivative  ${\cal F}''(1)$ at the critical point: ${\cal F}''(1){<}0$ on the left and ${\cal F}''(1){>}0$ on the right. The arrows show the effect of a small perturbation to the fixed points.
}
\label{f3}       
\end{figure}

Letting $\delta{=}\Pi-\Pi_{\text{c}}$ be a reduced control parameter, for $\delta, \Delta \ll 1$ the collapse process is described by the following differential equation:
\begin{equation}
\frac{d \zeta}{d t} \cong  \Delta +  A \zeta + B \zeta^2,
\label{2700}
\end{equation}
where $A{=}(1{-}\Delta)[1{+}(m{-}1)\delta]{-}1$ and $B=-(1{-}\Delta) {\cal F}''(1)$,
and we replaced ${\cal F}'(1){=}(m{-}1)P(1){=}1{+}(m{-}1)\delta$.
If the right-hand side of Eq.~(\ref{2700}) is zero for some $\zeta \ll 1$ the sequence of failures will stop cascading at that point, and the system is stable, suffering only a small-scale collapse.
On the other hand, if the right-hand side never becomes zero, the cascade of failures will produce a large-scale collapse of the unstable system (typically the whole system will fail).
Then, for small $\Delta$, whether the system undergoes a transition from stable to unstable when $\delta$ crosses from negative to positive values depends on the sign of $B \cong -{\cal F}''(1)$, as illustrated in Fig.~\ref{f3}.
Indeed, the discontinuous transition only occurs when the second derivative 
\begin{equation}
{\cal F}''(1) = \frac{m{-}2}{2(m{-}1)}- (m{-}1)P(2) 
\label{2800}
\end{equation}
 is negative at the critical point $\Pi_\text{c}=1/(m{-}1)$. This may be rewritten as the condition
 \begin{equation}
 P(2)  > \frac{m{-}2}{2(m{-}1)^2}
\label{2801}
\end{equation}
which becomes easier to satisfy the larger the number of layers.
 When ${\cal F}''(1)$ is positive, there is no collapse, and after damage the system returns to a stable configuration very close to $S=1$. If sufficient damage is applied, however, one may provoke the usual hybrid transition, see Fig.~\ref{f3}. 

\begin{figure}[h]
\begin{center}
\includegraphics[scale=0.65]{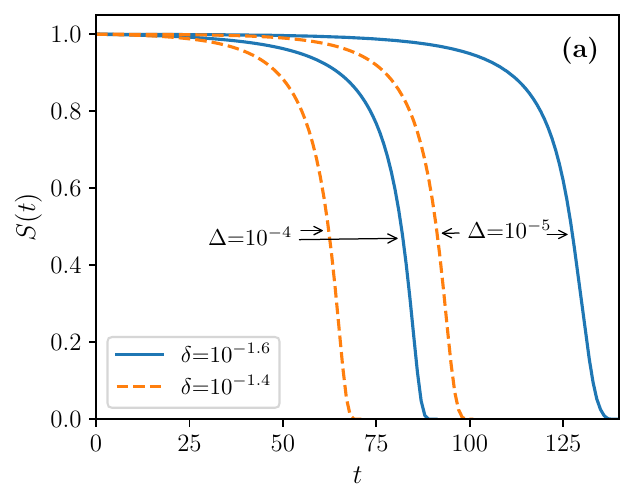}
\includegraphics[scale=0.65]{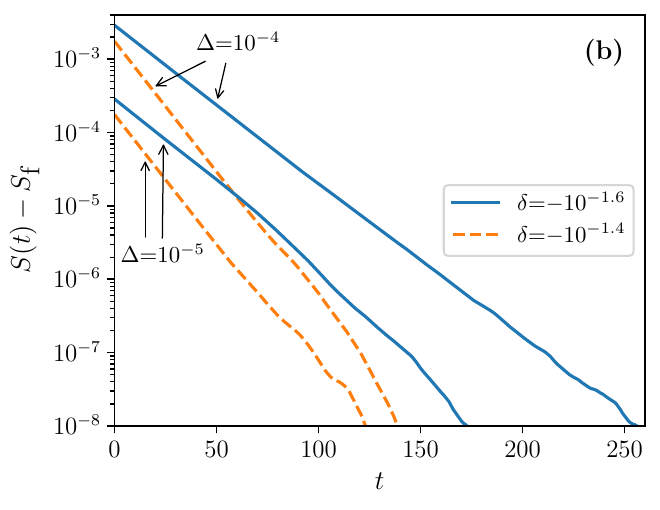}
\end{center}
\caption{
Evolution of $S(t)$ following the removal of edges with probability $\Delta$ in simulations of uncorrelated networks with $N=10^8$ and $m=3$ layers of edges, in the case $\Delta \ll \delta ^2$. 
The degree distribution is $P(0){=}0$ and $P(q){=}e^{-c}c^{q-1}/(q{-}1)!$ for $q{>}0$, with $c$ such that $\Pi \equiv P(1){=}e^{-c}{=}1/2 {+} \delta$.
(a) For $\Pi {>}\Pi_\text{c}$ ($\delta{>}0$) the network fully collapses.
(b) For $\Pi{<}\Pi_\text{c}$ and small damage $\Delta$ the fraction of surviving nodes $S(t)$ exponentially relaxes to some final $S_\text{f} {\sim} 1$.
These curves were obtained by averaging over $10$ realizations of the network and damage in (a) and over $100$ realizations in (b).
Obtaining smooth curves for the exponential relaxation cases requires a larger number of realizations because, due to the smallness of the variations of $S(t)$, fluctuations have a larger relative effect.
}
\label{fig_St}       
\end{figure}

\begin{figure}[h]
\begin{center}
\includegraphics[scale=0.65]{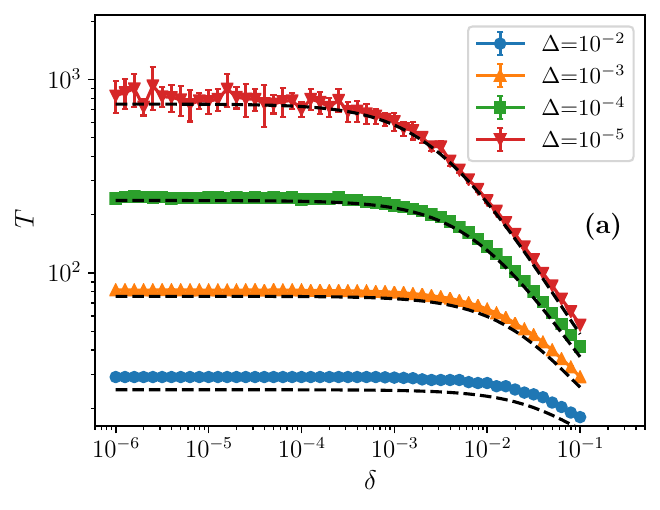}
\includegraphics[scale=0.65]{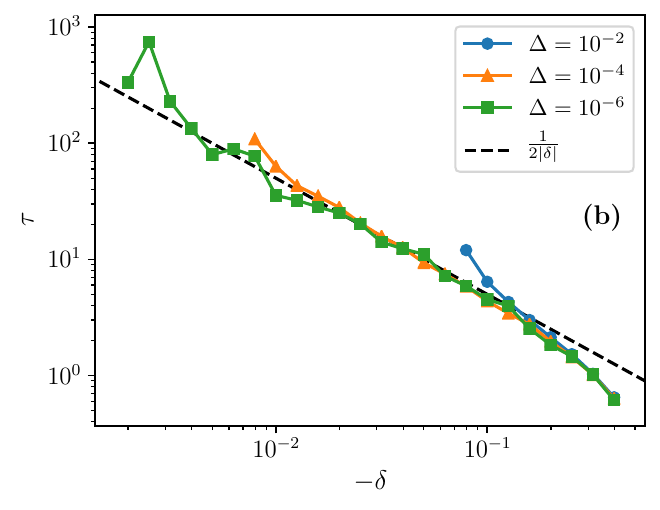}
\end{center}
\caption{
Dynamics of the collapse in simulations of uncorrelated networks with $N=10^8$ and $m=3$ layers as a function of $\delta$ for a range of values of $\Delta$.
The degree distribution distribution is $P(0){=}0$ and $P(q){=}e^{-c}c^{q-1}/(q{-}1)!$ for $q{>}0$, with $c$ such that $P(1){=}e^{-c}{=}1/2 {+} \delta$. 
Notice that for this $P(q)$ we have ${\cal F}''(1)<0$ at the critical point.
(a) Total collapse time $T$ for positive $\delta$; dashed lines represent Eq.~(\ref{as8}). 
(b) Relaxation rate $\tau$ for negative $\delta$. 
Each point is averaged over $10$ realizations of the network and damage in (a) and over $100$ realizations in (b).
In panel (b) points for $\Delta=10^{-6}$ are considerably noisier because the absolute number of edges removed, $\sim \Delta N = 10^2$, is already quite small.
}
\label{T_tau}       
\end{figure}

In the rest of this section we focus on how quickly the damage propagates through the network.
Namely, for unstable systems we calculate the number of pruning steps, $T$, necessary to collapse the whole network, while for stable systems we calculate the rate $\tau$ of the exponential relaxation law
\begin{equation}
 \zeta_\infty - \zeta   \propto e^{-t/\tau},
\label{2900}
\end{equation}
that describes the approach to the final state $\zeta_\infty \ll 1$.
In Fig.~\ref{fig_St}, we present the evolution of the fraction of surviving nodes at each step of the pruning process, $S(t)$, in simulations of a few representative cases, above and below $\Pi_\text{c}$.
For $\Pi{>}\Pi_{\text{c}}$ the system collapses completely, in a finite number of steps, while for $\Pi{<}\Pi_\text{c}$ the collapse is small-scale and $S(t)$ exponentially relaxes to some final $S_\text{f} {\sim} 1$.

\begin{table}[h]
  \begin{center}
    \caption{Asymptotics of collapse times for $\Delta ,\delta \to 0$.
    Expressions for $T$ represent total cascading time of unstable systems that suffer large-scale collapse, and expressions for $\tau$ represent the exponential relaxation rate of stable systems that suffer small-scale collapse.}
    \label{t3}
    \begin{tabular}{cccc} \hline\hline\noalign{\vskip 5pt} 
    & \multicolumn{2}{c}{$\Delta\ll\delta^2$}  & \multirow{2}{*}[-4pt]{$\delta^2 \ll\Delta$}
    \\
    \noalign{\vskip 5pt} 
    \cline{2-3}\noalign{\vskip 2pt} 
        & $\delta<0$ &  $\delta>0$ & 
    \\ \noalign{\vskip 2pt} \hline\noalign{\vskip 5pt} 
      ${\cal F}''(1)<0$ \, & {$\tau \cong \dfrac{1}{(m{-}1)|\delta|}$} \,& 
      $T \cong \dfrac{1}{ (m{-}1)\delta} \left\{
 \ln
 \left[\frac{(m-1)^2}{B}\right] 
 + 
 \left(\frac{\delta^2}{\Delta}\right)
 \right\}$
      & \,$T \cong \dfrac{\pi}{2\sqrt{B\Delta}}$
      \\
      \noalign{\vskip 5pt} 
      \hline
      \noalign{\vskip 5pt} 
      ${\cal F}''(1)>0$ & {$\tau \cong \dfrac{1}{(m{-}1)|\delta|}$} \,\,& $\tau \cong \dfrac{1}{(m{-}1)\delta}$  &  \,$\tau \cong \dfrac{1}{2\sqrt{-B\Delta}}$
      \\[10pt]
      \hline\hline
    \end{tabular}
  \end{center}
\end{table}

Depending on $\Delta$, $\delta$, and ${\cal F}''(1)$ not only the system may be stable or unstable, but also the relaxation rates and total collapse times, respectively, may be quite different. 
The form of the asymptotics $\Delta,|\delta|\ll 1$ is determined not only by the signs of $\delta$ and ${\cal F}''(1)$, but also by the relation between the magnitudes of $\Delta$ and $\delta^2$.
Detailed derivations of these results are given in Methods.

When ${\cal F}''(1)$ is negative, the transition exists. 
When $\delta > 0$, the system collapses when damaged. The time to collapse, $T$, is given by 
\begin{equation}
T \cong \frac{1}{ \sqrt{A^2{-}4B\Delta}} \ln \left[ \frac{A+\sqrt{A^2{-}4B\Delta}}{A-\sqrt{A^2{-}4B\Delta}} \right].
\label{as8}
\end{equation}
This theoretical result is compared with simulations in Fig.~\ref{T_tau}(a). As can be seen from the figure, there are two main asymptotic regimes, which may be obtained from Eq. (\ref{as8}) by considering either $\Delta \ll \delta^2$ or $\delta^2 \ll \Delta$. The expressions obtained are summarised in Table~\ref{t3}.
For very small perturbations in relation to the square of the distance from the critical threshold ($\Delta \ll \delta^2$), above the transition threshold ($\delta > 0$) the system collapses in a finite number of steps, which diverges as $1/\delta$, however some logarithmic dependence on $\Delta$ is retained.

Below the transition threshold, ($\delta < 0$), there is an exponential relaxation to a stable giant component, with relaxation time given by
\begin{equation}
\tau {\cong}  \left[A^2 {-} 4B\Delta \right]^{-1/2},
\end{equation}
see Methods.
In the regime $\Delta \ll \delta^2$, $\tau$ is asymptotically proportional to $1/|\delta|$. We see in Figs.~\ref{fig_St}(b) and \ref{T_tau}(b) that the relaxation rate is different for different values of $\delta$, but is independent of $\Delta$ in this regime.
These singularities $T,\tau \propto 1/|\delta|$ are distinct from the inverse square root singularity for $T$ and $\tau$ for a standard hybrid phase transition \cite{zhou2014simultaneous,baxter2015critical}.

When the square of the distance from the critical point is much smaller than the size of the perturbation ($\delta^2 \ll \Delta$), the time for the collapse $T$ diverges as $~1/\sqrt{\Delta}$, like in standard hybrid transitions.
In other words, for any finite $\Delta$, when $\delta$ is sufficiently small, the dependence on it disappears, so effectively there is no transition at $\delta=0$.
But in the limit of infinitesimal damage, i.e. $ \Delta \ll \delta^2 $, a hidden discontinuous transition emerges with non-standard exponent $T,\tau\propto |\delta|^{-1}$ (disregarding logarithmic corrections).
The derivation of Eq. (\ref{as8}) and the asymptotic expressions are given in Methods.

When ${\cal F}''(1)$ is positive, there is no collapse, but interestingly there is a very close relationship between the expressions for the relaxation time $\tau$ and those described above for $T$, having the same dependence on $\delta$ or $\Delta$ in almost all cases, see Table~\ref{t3} and Methods.


\section*{Discussion}
\label{concl}

In this paper we have shown that the absence of bare nodes [$P(0)=0$] produces an exotic hidden transition in a reference percolation problem for multiplex networks, weak multiplex percolation. 
The transition, a discontinuity without any singularity of the size of the giant component, can be observed by removing a vanishingly small fraction of vertices or edges from the network. 
The fraction of nodes of degree $1$ in each layer can be used as the control parameter, $\Pi \equiv P(1)$.
Above $\Pi=\Pi_\text{c}$, this removal destroys the giant component completely, while below $\Pi_\text{c}$, the removal has a negligible effect on the giant component.

In symmetric, uncorrelated multiplex networks with $m$ layers, the size of the weakly percolating giant component can be calculated by solving self consistency equations for the probability $Z$ that following a random edge in one layer leads to a node which has at least one edge in every other layer satisfying the same condition.
For the hidden transition to appear, the equation for $Z$ must have the fixed point $Z=1$, and this point must be stable on one side from the critical point and unstable on the other. 
We show that, in this problem, 
(i) when $m=2$, this fixed point is always stable and the hidden transition is absent; 
(ii) when $m \geq 3$, the hidden transition occurs at $\Pi = 1/(m{-}1)$ for uncorrelated networks. 
The transition also appears in correlated networks. For example, in networks with ultimately correlated degree distributions one may observe the same kind of discontinuous transitions, with only the position of the critical point changing to $\Pi_\text{c}=\langle q \rangle/(m{-}1)$. However the transition is not possible in the alternative generalization of percolation to multiplex networks, the mutually connected component.

To understand the dynamics of the collapse, we expand the self consistency equations in terms of the amount $\Delta$ of random damage. 
Although the mechanism of collapse in this hidden transition is essentially the same as that of the hybrid transition, the singularities of the relaxation time and time to collapse are very different.
We show that the collapse above $\Pi_\text{c}$ takes a finite time $T \propto [\Pi{-}\Pi_\text{c}]^{-1}$. 
There is an exponential relaxation below $\Pi_\text{c}$ with a relaxation time $\tau \propto [\Pi_\text{c}-\Pi]^{-1}$. 
The exponent for the relaxation time equals $1$ in contrast to a hybrid transition, where it equals $1/2$.


\section*{Methods}

\subsection*{Uncorrelated symmetric layers}
\label{app0}

Let us consider the case in which each layer has the same degree distribution $P(q)$, with no correlations between layers. Since the degree distribution is symmetric across layers, the probability $Z_\alpha$ is the same for all $\alpha$, so (dropping the subscript) $Z$ obeys Eq. (\ref{2300}):
\begin{equation}
Z = \left\{\sum_q P(q) \left[1-(1-Z)^q\right] \right\}^{m-1} \equiv {\cal F} (Z).
\label{Z_a0}
\end{equation}
The only nodes that meet the criterion for removal are the ones without any connections, so if $P(0)=0$, the pruning procedure will not remove any of the edges in the original graph. In other words, in this case, $Z=1$ is a fixed point of this equation. 

\begin{figure}[t]
\begin{center}
\includegraphics[scale=0.75]{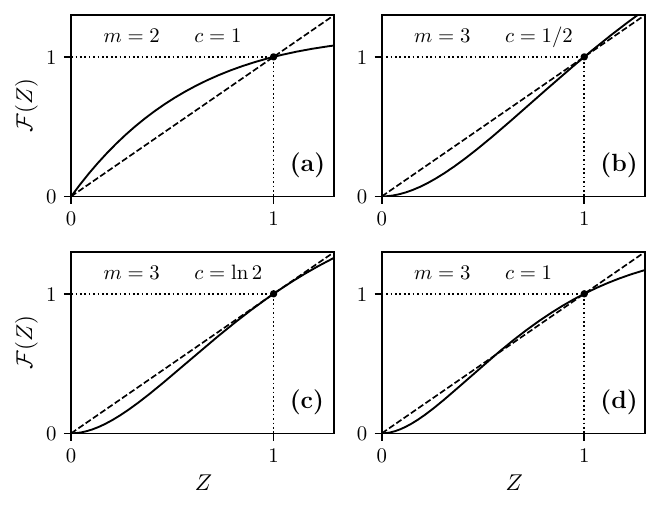}
\end{center}
\caption{
Function ${\cal F}(Z)$
for symmetric uncorrelated offset Poisson degree distribution, $P(q)=e^{-c}c^{q-1}/(q-1)!$ for $q{>}0$, and
$P(0){=}0$. 
The network is undamaged, and $Z=1$ is always a fixed point.
(a) In $m=2$ layers the fixed point is stable for any $c>0$.
In $m=3$ layers the stability transition happens at $P(1)=1/2$, corresponding to $c = \ln 2$. Panel (b) shows a profile of $ {\cal F}(Z)$ in the unstable phase for $c<\ln 2$, (c) represents the critical point $c=\ln 2$, and (d) the stable phase for $c>\ln 2$.
}
\label{f4}       
\end{figure}

Let us expand ${\cal F}(Z)$ near the fixed point $Z=1$,
\begin{equation}
{\cal F}(Z) = 1-{\cal F}'(1) [1{-}Z(t)]+{\cal F}''(1) [1{-}Z(t)]^2+...
\label{60}
\end{equation}
The coefficient of the second term is ${\cal F}'(1) = (m-1) P(1)$. When this coefficient is larger than $1$ the fixed point is unstable, otherwise it is stable.
For $m=2$ the derivative ${\cal F'}(1)$ is never larger than $1$ because $P(1) < 1$, and $Z=1$ is a stable fixed point.
In the case of $m>2$ layers ${\cal F'}(1)$ can be smaller or larger than $1$, and thus, the fixed point can be stable or unstable, respectively, see Fig.~\ref{f4}.
The stability transition occurs when the $P(1)={1}/(m{-}1)$.


For two layers we have
\begin{equation}
Z = \sum_q P(q) \left[1-(1-Z)^q\right],
\label{70}
\end{equation}
which has two fixed points, $Z=0$ and $Z=1$, see Fig.~\ref{f4}. 
The right-hand side of Eq.~(\ref{70}) is always above $Z$ in the interval $0<Z<1$. 
So, the fixed point $Z=0$ is unstable and the fixed point $Z=1$ is stable, which means that a small perturbation will not collapse the system. 
Thus, after the application of an infinitesimal amount of damage, we still have 
$S = 1 
.
$


For three layers we have
\begin{equation}
Z = \left\{\sum_q P(q) \left[1-(1-Z)^q\right] \right\}^{2} .
\end{equation}

The fixed points generally have the same behavior for any reasonable degree distribution $P(q)$. 
%
For illustrative purposes, it will be useful to refer to a concrete example. Consider an offset Poisson degree distribution $P(q) = c^{q-1} e^{-c} / (q-1) !$. Then Eq. (\ref{Z_a0}) becomes
\begin{equation}
Z = {\cal F}(Z) = \left[1 - (1-Z)e^{-cZ}\right]^{m-1}
\label{Z_pois}
\end{equation}
and
\begin{equation}
S = \left[1 - (1-Z)e^{-cZ}\right]^{m}\,.
\end{equation}

 In three layers this becomes
\begin{equation}
Z = 1 - 2(1-Z)e^{-cZ}+  (1-Z)^2e^{-2cZ} = {\cal F}(Z) \,.
\label{100}
\end{equation}
This function is plotted in Fig.~\ref{f4}.
In the interval $0\leq Z \leq 1$, 

(i) If $c<\ln 2$, it has two fixed points, $Z=0$ and $Z=1$, and  
${\cal F}(Z)$  is below $Z$ in the interval $0<Z<1$, so the fixed point $Z=0$ is stable and the fixed point $Z=1$ is unstable. Thus $S(c<\ln 2)=0$. 

(ii) At $c=\ln 2$, the derivative of ${\cal F}(Z)$ equals $1$ at $Z=1$. 

(iii) If $c>\ln 2$, Eq.~(\ref{100}) has three fixed points (within the interval $0 {\leq} Z {\leq} 1$): $Z=0$, $Z=1$, and 
the third fixed point $0{<}Z_3{<}1$. 
(As $c$ increases from $\ln 2 +$ to $\infty$, $Z_3$ decreases from $1-$ to $0$.) 
The function ${\cal F}(Z)$ is less than $Z$ in the interval $0{<}Z{<}Z_3$.  
It is greater than $Z$ in the interval $Z_3{<}Z{<}1$.  
So the point $Z_3$ is unstable and the points $0$ and $1$ are stable. 

Thus, we have a transition at 
$c_\text{c}=\ln 2
$. 
Below this transition any perturbation will result in $S=0$. 
Above this transition the system will be essentially unchanged by a small amount of damage,  
$S = 1$.
%
The critical point depends on the degree distribution, in particular, on $P(1)$, but this qualitative picture applies to any distribution.

For $m>3$ the structure of the equations is the same as for $m=3$, and the results are qualitatively similar. 
For the offset Poisson distribution, for example, the critical point $c_\text{c}=\ln (m{-}1)$.
For $c>c_\text{c}$, we have $S = 1$. 
Thus we have the phase transition 
\begin{equation}
S =  \theta[c - \ln (m{-}1)]
.
\label{1000}
\end{equation}
where $\theta$ is the Heaviside function.

\subsection*{Ultimate Correlations}
\label{app1}

Let us consider the extreme form of assortative correlations where nodes have the same degree in all layers, that is
\begin{equation}
P(q_1, ... ,q_m) = P(q) \prod_{\alpha=1}^m \delta_{q_\alpha, q}
,
\label{31}
\end{equation}
where $ \delta_{q_\alpha, q}$ is the Kronecker delta function. We show that the qualitative behavior is the same as for the uncorrelated case.

Since the degree distribution is symmetric across layers, the probability $Z_\alpha$ is same for all $\alpha$, and Eq.~(\ref{10}) gets simplified to
\begin{equation}
Z = \sum_q \frac{q P(q)}{\langle q \rangle} \bigl[ 1 - (1-Z)^q \bigr]^{m-1} = {\cal F}(Z)
.
\label{41}
\end{equation}

In the case of ultimate correlations, the value $Z=1$ is a fixed point of Eq.~(\ref{41}) regardless of $P(q)$ (including when $P(0)>0$), because no edge will be connected to a node with degree zero in another layer.
The expansion of ${\cal F}(Z)$ becomes
\begin{equation}
{\cal F}(Z) = 1 - \frac{(m{-}1) P(1)}{\langle q \rangle} (1{-}Z) + ...
.
\label{61}
\end{equation}

The value of $S$ at the fixed point $Z=1$ is different from $1$, however. Eq.~(\ref{20}) becomes
\begin{equation}
S = \sum_q P(q) \bigl[ 1 - (1-Z)^q \bigr]^m
.
\label{51}
\end{equation}
So at $Z=1$ we have
$S = 1 - P(0)\,$.

For $m=2$ the derivative ${\cal F'}(1)$ again is never larger than $1$ because $\langle q \rangle {\geq} P(1)$, and $Z=1$ is stable fixed point.

In the case of $m>2$ layers, just as in the uncorrelated symmetric case, ${\cal F'}(1)$ can be smaller or larger than $1$, and thus, the fixed point can be stable or unstable, respectively, see Fig.~\ref{f4}.
The stability transition occurs when the $P(1)={\langle q \rangle}/(m{-}1)$.
Above this transition threshold any perturbation will result in $S=0$. 
Below this point, the system will be essentially unchanged by a small amount of damage. 
Thus we have the phase transition 
\begin{equation}
S = [1 - P(0)]\, \theta[\langle q\rangle - (m{-}1)P(1)]
.
\label{1001}
\end{equation}


\subsection*{Asymptotics of Collapse Times}
\label{app2}

Here we derive the asymptotics of small- and large-scale collapse times of uncorrelated networks in terms of the damage probability $\Delta$, the deviation to the critical point $\delta=\Pi-\Pi_\text{c}$, and the second derivative of the function ${\cal F}(Z)$ at the fixed point $Z=1$.
There are two definitions for these times: while, for large-scale collapse, $T$ is the number of pruning iterations that unstable networks take to collapse completely, for small-scale collapse, the characteristic time $\tau$ is the relaxation rate of the exponential convergence to the stable fixed point $\zeta_\infty\ll1$ in Eq.~(\ref{2900}).

For simplicity, let us rewrite Eq.~(\ref{2700}) as
\begin{equation}
\frac{d \zeta}{d t} =  \Delta + A \zeta+ B \zeta^2 + ... , 
\label{as2}
\end{equation}
where $A{=}(1{-}\Delta)[1{+}(m{-}1)\delta]-1$, and $B{=}{-}(1{-}\Delta){\cal F}''(1)$.
Letting $\delta, \Delta \ll 1$ we have 
\begin{equation}
A \cong (m{-}1)\delta - \Delta
\end{equation}
and
\begin{equation}
B \cong - {\cal F}''(1) = (m-1)P(2) {-} \frac{m{-}2}{2(m{-}1)}\,.
\end{equation}

Assuming $B \neq 0 $, for a solution $\zeta_\infty$ to exist, the polynomial $\Delta + A \zeta+ B \zeta^2$ must have at least one root that is a real number larger than zero.
There are four different possibilities:

(i) When $B<0$, the two roots are always real but only the largest is positive. 

(ii) When $0<B<\frac{A^2}{4\Delta}$ and $A<0$, both roots are positive, and the physical solution $\zeta_\infty$ corresponds to the smallest of them. 

(iii) When $0<B<\frac{A^2}{4\Delta}$ and $A>0$, both roots are negative, and the system collapses to $\zeta_\infty$ far away from zero, 'usually' to $\zeta_\infty=1$. 

(iv) When $B>\frac{A^2}{4\Delta}$, the roots are never real numbers, and the system also collapses to $\zeta_\infty$ far away from zero.

\subsubsection*{Exponential relaxation of small-scale collapse}

Cases (i) and (ii) result in an exponential relaxation to a fixed $\zeta_\infty$ close to zero that is given by the following expression in both cases,
\begin{equation}
 \zeta_\infty = -\frac{A{+}\sqrt{A^2 {-} 4B\Delta}}{2B}.
\label{as3}
\end{equation}
The exponential approach to this point is calculated by expanding the right-hand side of Eq.~(\ref{as2}) near $\zeta_\infty$,
\begin{equation}
\frac{d (\zeta_\infty {-} \zeta )}{d t}  = {-} \sqrt{A^2 {-} 4B\Delta} (\zeta_\infty{-} \zeta) + ..., 
\nonumber
\end{equation}
which immediately gives
\begin{equation}
 \zeta_\infty - \zeta   \propto e^{-t/\tau},
\label{as4}
\end{equation}
where $\tau {\cong}  \left[A^2 {-} 4B\Delta \right]^{-1/2} $.


For $\delta < 0$ we always have $A < 0$, therefore the exponential relaxation occurs for all $B<\frac{A^2}{4\Delta}$.
Specifically, on one hand, for $\Delta \ll \delta^2$ we have 
\begin{equation}
\tau {\cong} \frac{1}{(m-1)|\delta|} ,
\label{as5}
\end{equation}
even for large positive values of $B$, which intuitively aligns with the independence of this result from $B$. 
In this case we can even get the prefactor of the exponential of Eq.~(\ref{as4}), which is simply $ \frac{\Delta}{(m{-}1)|\delta|}$ and coincides with $\zeta_\infty$.

On the other hand, when $ \delta^2 \ll \Delta$ we have 
\begin{equation}
\tau {\cong} \frac{1}{2\sqrt{-B\Delta}}. 
\label{as6}
\end{equation}
Which applies for all $B < 0$ except for a vanishing region close to $B=0$ whose size is of order $\Delta$, and which includes the positive values of $B$ up to the maximum value $ \frac{A^2}{4\Delta}$.


When $\delta > 0$ the situation is somewhat different for $\Delta \ll \delta^2$.
Namely, $A>0$, so we must have $B<0$ for exponential relaxation to occur, leading to an expression symmetric to Eq.~(\ref{as5}):
\begin{equation}
\tau {\cong} \frac{1}{(m-1)\delta} ,
\label{as51}
\end{equation}
except that here $B$ cannot be positive. 
Remarkably, however, as long as $B < 0$, the relaxation rate $\tau$ does not depend on its particular value.

If $\delta^2 \ll \Delta$ then $A$ may be positive or negative. 
Nevertheless, both cases lead back to Eq.~(\ref{as6}) with the same condition $ B < 0$  for exponential relaxation to occur.

\subsubsection*{Large-scale collapse in finite time}

Large-scale collapse happens in the cases (iii) and (iv), which cover all the situations where the polynomial  $\Delta + A \zeta+ B \zeta^2$ does not become zero for some small positive $\zeta_\infty$.
In these cases the system collapses in a finite time $T$, which can be estimated by rearranging and integrating Eq.~(\ref{as2}) as follows
\begin{equation}
T \cong \int_0^\infty \frac {1}{ \Delta + A \zeta+ B \zeta^2 } d \zeta.
\label{as7}
\end{equation}
To be precise, we should write the upper limit of the integral as $1$.
However, since we are looking at the limit of small $\Delta$ and $\delta$, the contribution from the upper limit can be neglected, because the contribution of the lower limit is much larger.
This is a consequence of most of the pruning steps being spent escaping from the small $\zeta$ region.

Then, total collapse time is given by the formula
\begin{equation}
T \cong \frac{1}{ \sqrt{A^2{-}4B\Delta}} \ln \left[ \frac{A+\sqrt{A^2{-}4B\Delta}}{A-\sqrt{A^2{-}4B\Delta}} \right],
\label{as8app}
\end{equation}
which is Eq. (\ref{as8}) of the main text.
In the case that the square root is complex, one can instead write this expression in terms of $\arctan$ functions involving only real numbers.

Since $B$ is positive (i.e., ${\cal F}''(1)<0$), a small amount of damage $\Delta$ will always cause the total collapse if $\delta^2 \ll \Delta$, for $\delta$ both above and below the critical point $\delta=0$.
The asymptotics of Eq.~(\ref{as8}) this case are 
\begin{equation}
T \cong \frac{\pi}{ 2 \sqrt{B\Delta}} \cong  \frac{\pi}{ 2 \sqrt{|{\cal F}''(1)|\Delta}}.
\label{as9}
\end{equation}

Complementarily, in the limit of infinitesimal amount of damage $\Delta \to 0$, we are in the region $\Delta \ll \delta^2$, which leads to two different situations below and above the transition.
On one hand, when $\delta{<}0$ the fixed point $Z=1$ is stable, and the system undergoes a small-scale collapse as described in the previous section.
On the other hand, when $\delta{>}0$ the fixed point $Z=1$ is unstable, and the system collapses completely (even for the tiniest amount damage) in a finite time,
\begin{equation}
T \cong \frac{1}{ (m{-}1)\delta} \left\{
 \ln\left[\frac{(m-1)^2}{B}\right] + \ln\left(\frac{\delta^2}{\Delta}\right)
\right\}\,.
\label{as10}
\end{equation}


\bibliography{weak_hidden_refs}

\section*{Acknowledgements}

This work was developed within the scope of the project i3N, UIDB/50025/2020 \& UIDP/50025/2020, financed by national funds through the FCT/MEC. This work was also supported by National Funds through FCT, I. P. Project No. IF/00726/2015. R. A. d. C. acknowledges the FCT Grants No. SFRH/BPD/123077/2016 and No. CEECIND/04697/2017.

\section*{Author contributions statement}

The study was designed by S.D. and R.C. with input from G.B. and J. M. Analytical calculations were carried out by S.D. and R.C. and reviewed by G.B. The manuscript was written by R.C. and G.B. Numerical calculations and production of figures were carried out by R.C. All authors contributed to the planning and revision of the manuscript.

\end{document}